# Comprendre l'origine du réchauffement climatique
## Comment construire le lien entre émissions de CO$_2$ et température moyenne de la Terre ?

Maron Valentin[1][2]
Dufresne Jean-Louis[3]
Pélissier Lionel[1][2]
Rabier Alain[2]
Cochepin Medhi[2]
[1]EFTS (Education, Formation, Travail, Savoirs), Université Toulouse Jean Jaurès – France
[2]INSPE Toulouse Occitanie-Pyrénées, Université Toulouse Jean Jaurès – France
[3]IPSL (Institut Pierre-Simon-Laplace), École normale supérieure - Paris, Université Paris Cité – France

## Résumé

Nous proposons dans ce travail une approche d'enseignement du réchauffement climatique adaptée aux nouveaux programmes de lycée général et professionnel. L'objectif visé est la construction du lien entre émissions de CO$_2$ et élévation de la température moyenne de la Terre, en s'appuyant autant que possible sur des éléments empiriques explicites. Nous présentons dans un premier temps les approches d'enseignement existantes sur le sujet, qui sont ensuite mises en relation avec les idées et modes raisonnement connus des élèves, en particulier : la non évidence des interactions entre gaz et rayonnement pour les élèves. Les choix didactiques structurant l'approche proposée sont alors exposés. Ils s'articulent autour de l'interprétation d'expériences réalisées avec une caméra infrarouge d'une gamme spectrale permettant de voir l'émission et l'absorption du CO$_2$.

## Contexte

« Le réchauffement climatique annonce des conflits comme la nuée porte l'orage. [...] ce qui est en cause avec cette conférence sur le climat, c'est la paix. ». Ces mots lors du discours d'introduction de la COP21 en 2015[1], par le président de la République française, rappelle tout l'enjeu de l'action climatique. Une des conditions nécessaires pour que des choix politiques forts puissent être acceptés par la population est qu'une majorité suffisante soit d'accord sur l'origine du problème. S'il est clair dans le dernier rapport du GIEC[2] que *l'intégralité* du changement climatique observé actuellement est dues aux activités humaines, cela ne fait pas encore consensus dans la population. En 2022, d'après une enquête EDF-IPSOS[3], 37% des sondés (24000 personnes dans 30 pays) désapprouvent l'idée que « le changement climatique actuel est principalement due à l'activité humaine ». Une proportion en hausse de 6% relativement à 2019. Ce pourcentage atteint 48% aux Etats-Unis, en Russie et en Norvège, ou encore 60% en Arabie Saoudite, les quatre premiers pays premiers producteurs d'énergies fossiles. Parmi les raisons de ce constat figurent les stratégies de désinformations massives de l'industrie fossiles aujourd'hui bien connues (Oreskes & Conway, 2010). Ces phénomènes sont également présents en France, avec en particulier un fort regain depuis l'été 2022 sur les réseaux sociaux (Chavalarias et al., 2023). Celui-ci peut être relié à la hausse de 8% en France des personnes niant l'origine principalement humaine du changement climatique, entre 2019 et 2022[3]. Face à cette situation, le rôle de l'enseignement pour établir cette connaissance de la manière la plus convaincante possible est crucial. En France comme dans d'autres pays, la place de la physique du climat a été fortement renforcée dans les nouveaux programmes du secondaire, en particulier dans le cadre de l'Enseignement Scientifique des classes de Première et Terminale générale, ainsi que dans le programme de physique du lycée professionnel. Ces ajouts étant relativement récents, les ressources éducatives spécifiques à ces programmes sont peu nombreuses, de même que les recherches en didactique de la physique sur le sujet.

Ce travail, issu d'une collaboration entre didacticiens et physiciens du climat, a pour objectif de développer une approche d'enseignement visant à construire le lien entre émissions de dioxyde de carbone ($CO_2$) et élévation de la température moyenne de la Terre. La mise en évidence de ce lien apporte un premier argument qualitatif pour comprendre l'origine humaine du changement climatique, en amont de la modélisation permettant les simulations du climat passé et futur.

---

[1] https://www.elysee.fr/front/pdf/elysee-module-13543-fr.pdf

[2] https://www.ipcc.ch/report/ar6/wg1/#SPM

[3] https://www.edf.fr/groupe-edf/observatoire-international-climat-et-opinions-publiques/telechargements

Une intention centrale de cette approche est qu'elle soit accessible avec le minimum de prérequis conceptuel, afin d'être utilisable pour des élèves non spécialistes de physique, notamment dans le cadre de l'Enseignement Scientifique, adressé à l'ensemble des élèves de lycée. Cette contrainte permet également de rendre cette approche potentiellement utilisable pour la vulgarisation scientifique. Notre hypothèse est qu'une compréhension approfondie du lien entre $CO_2$ et température peut servir de point d'appui à d'autres modalités de l'éducation au changement climatique.

## Cadre méthodologique

Ce travail se situe dans le cadre du « *Model of Educational Reconstruction* » (Kattmann et al., 1996), basé sur la mise en relation de l'analyse du contenu scientifique en jeu avec l'analyse des points de vue des élèves sur le sujet, dans le but de définir les éléments de contenu clefs pour l'enseignement à un niveau donné. L'objectif final est de produire une « *structure open source* », au sens de (Besson, Borghi, et al., 2010) :

> *To bridge the gap between research project and school reality and, to facilitate the reproducibility in an actual classroom context, the sequences are designed in a form that we call an 'open-source structure'. This means that there is a core of contents, conceptual correlations and methodological choices with a cloud of elements that can be re-designed, omitted, or added by the teacher, who may thereby create new versions of the sequence. Some elements redesigned by the teachers can be included in the standard version issued by the research group. In this sense, the teachers' work provides useful feedback that allows us not only to test the effectiveness of the proposal and to identify its points of weakness, but also to enrich it with new elements. A variety of approaches is expected in the actual organization of educational activities, depending on the constraints of the classroom and teacher preferences.* (Besson, Borghi, et al., 2010)

Nous commencerons par une présentation d'approches d'enseignement du sujet parmi celles les plus développées, que nous questionnerons ensuite relativement aux idées a priori et difficultés connues vis-à-vis de la compréhension de l'effet de serre. Cette mise en perspective conduira à expliciter nos questions de recherche, puis à présenter et justifier les différents choix didactiques constituant la proposition issue de ce travail.

## Analyse d'approches d'enseignement existantes

La première mise en évidence du rôle du $CO_2$ sur le climat a été établi par le scientifique Svante Arrhenius dans son article « *On the influence of carbonic acid in the air upon the temperature of the ground* » (1896). Depuis, les interactions entre l'atmosphère et le rayonnement sont compris de plus en plus précisément par les spécialistes du domaine et intégrées dans les modèles de climat. Les équations en jeu pour rendre compte de ces phénomènes sont accessibles à un niveau universitaire, comme par exemple dans le cours d'introduction à la physique de l'atmosphère[4] du magistère des Sciences de la Terre à Paris.

Parmi les scientifiques spécialistes du sujet, certains ont travaillés particulièrement sur des approches de vulgarisation du phénomène d'effet de serre atmosphérique, et ont produit des ressources pédagogiques destinées aux enseignants, dont notamment les deux suivantes, parmi les plus récentes et les plus développées :

- « Principes de base de l'effet de serre » (Dufresne, 2020)
- « Rayonnement, opacité et effet de serre » (Thollot & Dequincey, 2021)

Un point commun de ces deux ressources est le fait de montrer que l'absorption du rayonnement dépend de la nature du matériau et du type de rayonnement. Suite à ces expérimentations impliquant des solides, est amenée l'idée que cela est également valable pour les gaz. La référence empirique utilisée est le spectre d'absorption de différents gaz. Il s'agit également de l'approche choisie dans une récente recherche en didactique de la physique (Toffaletti et al., 2022).

Dans le cadre de ce travail, visant le développement d'une approche la plus accessible possible du sujet, ces choix didactiques ne sont pas adaptés. En effet, pour un public non spécialiste de physique, ayant une faible maîtrise (voir nulle) de la notion de longueur d'onde électromagnétique, les spectres d'absorption des gaz s'avèrent difficiles (voir impossibles) à comprendre, et donc à utiliser comme référence empirique.

Une autre approche courante dans les ressources pédagogiques sur le climat (par exemple sur le site de *La main à la pâte*[5]), ainsi que dans certaines recherches en didactique (Besson, De Ambrosis, et al., 2010), consiste à s'appuyer sur une manipulation en apparence simple. Celle-ci consiste à comparer l'élévation de température d'une enceinte fermée remplie de dioxyde de carbone et une autre remplie d'air, pareillement exposées au rayonnement. Si on constate effectivement une élévation de température plus grande dans le cas du $CO_2$, cela est dû principalement aux phénomènes de conduction et convection, variant avec la densité du gaz, et non à l'interaction du $CO_2$ avec le rayonnement infrarouge, comme dans le cas de l'atmosphère. Une démonstration de la fausseté du raisonnement est notamment faite dans l'article *« Climate change in a shoebox: Right result, wrong physics »* (Wagoner et al., 2010).

---

[4] https://www.lmd.jussieu.fr/~hourdin/PEDAGO/cours.pdf

[5] https://fondation-lamap.org/sequence-d-activites/co2-effet-de-serre-et-activites-humaines

Si cette expérience simple ne correspond pas au phénomène d'interaction entre gaz et rayonnement ayant lieu dans l'atmosphère, il ne reste donc aucune preuve expérimentale accessible du rôle du $CO_2$ dans le réchauffement climatique. L'unique point d'appui utilisable est l'analogie avec le verre ou le plastique, via les expériences montrant l'absorption du rayonnement infrarouge. D'un point de vue didactique, l'utilisation de cette analogie pose la question de sa recevabilité par les élèves, au-delà l'argument d'autorité.

## Analyse des idées a priori et difficulté en jeu

### A propos de l'interaction entre gaz et rayonnement

Des recherches ont montré qu'au niveau de l'école primaire, l'idée que « l'air soit de la matière » n'est pas du tout une évidence pour les élèves (Plé, 1997). D'un point de vue physique, les gaz ne partagent pas toutes les propriétés des solides. Pourquoi partageraient t'ils a priori leurs propriétés concernant les interactions avec le rayonnement ? Pour les élèves, dans quelle mesure l'extrapolation des solides au gaz est-elle intuitive ? Une enquête a été conçue afin d'avoir des éléments de réponse sur ce point, menée par des étudiants de master MEEF (Amiel & Aubry, 2023; Cochepin, 2022; Romond & Lusson, 2023). Suite à une introduction classique à l'émission et absorption du rayonnement infrarouge par des solides, il est demandé aux élèves leur avis a priori concernant les gaz. Les résultats sont quasiment les mêmes pour les deux questions (émission et absorption) : plus de 3/4 des collégiens (N=208) et 2/3 des lycéens (N=280) répondent soit que les gaz ne peuvent ni émettre ni absorber de rayonnement, soit qu'ils ne savent pas.

Face à ce constat, la question se pose de la manière d'introduire ce phénomène sans le donner à admettre, étant donné le caractère crucial de ce point pour faire le lien entre émissions de $CO_2$ et réchauffement climatique. A notre connaissance, aucun article de recherche en didactique de la physique, aucun manuel de Terminale Enseignement Scientifique, ni aucun site proposant des ressources pédagogiques ou de vulgarisation sur le climat (Eduscol[6], Office for Climate Education[7], EduClimat[8], Bonpote[9]) n'apporte d'élément empirique convaincant et accessible concernant les capacités d'émission et d'absorption du rayonnement par certains gaz.

### A propos de l'effet de serre

Une littérature importante existe à propos des difficultés relatives à la compréhension de l'effet de serre (Gautier et al., 2006; Shepardson et al., 2011; Handayani et al., 2021). Les points convergents de ces recherches sont synthétisées notamment dans (Toffaletti et al., 2022). Parmi ceux-ci revient de manière récurrente l'idée selon laquelle les gaz à effet de serre constituent une sorte de « couche » ou de « couvercle » dans l'atmosphère, formant ainsi une « barrière » contre laquelle une partie de la chaleur ou le rayonnement émis par la Terre « rebondirait » pour revenir vers elle, « piégeant » ainsi l'énergie du Soleil.

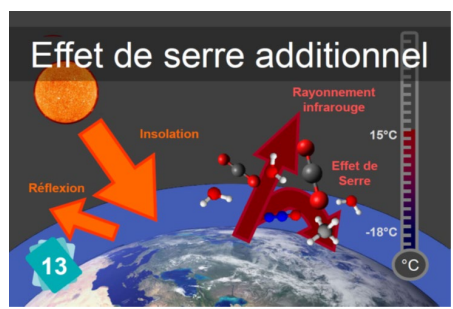

Figure n° 1 : schéma extrait de la Fresque du climat

Cette idée de « piégeage » est régulièrement présente dans les textes de vulgarisation et les manuels d'enseignement, et est renforcée par les schémas proposés (comme par exemple celui de la figure n°1). Cette notion implique implicitement une idée de chronologie, avec rayonnement qui part du sol, puis qui est empêché de sortir par l'atmosphère et qui revient ensuite vers la surface de la Terre, et contribue à la réchauffer « un peu plus ». Ce raisonnement de type séquentiel est incompatible avec les schémas associés la plupart du temps, représentant des bilans instantanés, c'est à dire où les valeurs des puissances de rayonnements sont considérées au même moment, et où la température est constante (Colin & Tran Tat, 2011). Un autre aspect problématique, relevé à la fois chez les élèves et dans les ressources de vulgarisation, est la non distinction entre les différents types de rayonnement (solaire, infrarouge terrestre, infrarouge atmosphérique). Cette confusion participe à rendre possible l'idée de « rebond » ou de « piégeage » du rayonnement venant du sol.

---

[6] https://eduscol.education.fr/1132/changement-climatique

[7] https://www.oce.global/fr

[8] https://educlimat.fr/

[9] https://bonpote.com/

## Problématique

Etant donné les différents problèmes d'enseignement et d'apprentissage présentés ci-dessus, la question centrale de cette recherche est la suivante : comment construire le lien entre émissions de $CO_2$ dans l'atmosphère et température moyenne de la Terre, pour un public non spécialiste de physique ? Plus précisément, comment établir cette relation en s'appuyant sur des éléments empiriques accessibles avec un minimum de prérequis, tout en prenant en compte les idées a priori sur le sujet ?

## Proposition de structure didactique pour relier $CO_2$ et température

### Introduction au rayonnement infrarouge

La première notion à construire dans l'approche proposée est celle de rayonnement infrarouge. Son introduction s'appuie sur sa détection par une caméra infrarouge (figure n°2). Le choix a été fait de faire cette image dans le noir, de manière à éliminer une possible confusion entre les deux manières pour un objet d'être visible : par diffusion (source secondaire) ou émission (source primaire).

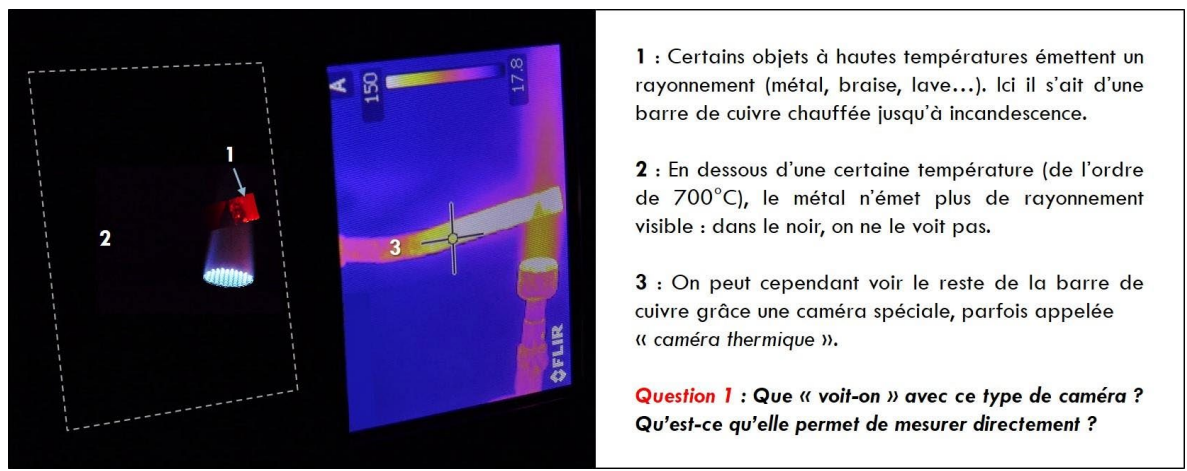

Figure n°2 – Une barre de cuivre chauffée jusqu'à incandescence, dans le noir, observée avec une caméra « thermique »

Le terme de « caméra thermique » est en soit problématique dans la mesure où il renforce l'idée d'une caméra permettant de « voir » la température, à la façon d'un thermomètre. Cela est d'autant plus renforcé par l'échelle de température (reconstruite) associée à ce type d'image, comme dans la figure n°2 (en haut à droite). La suite du cheminement nécessite la compréhension de ce que mesure directement ce type de caméra : des puissances de rayonnement infrarouge. Des premières expérimentations en classe sur ce point (Cochepin, 2022) ont montré la forte tendance à continuer d'interpréter les images infrarouges en termes de température ou de chaleur, malgré une clarification de ce point en début de cours. Cette conception erronée interfère par la suite avec l'interprétation des images mettant en jeu les phénomènes d'émissions et absorption du rayonnement infrarouge, où les couleurs ne correspondent plus aux températures des objets observés. Afin d'anticiper et d'insister sur ce point, il a été choisi de considérer la réflexion infrarouge (figure n°2), dans le but mettre en évidence l'impossibilité de l'interprétation en température, et le point commun avec la lumière visible. Dans toute la suite, il est intentionnel d'avoir laissé présent ce reflet infrarouge en arrière-plan des autres expériences, afin de pouvoir s'y référer régulièrement pour rappeler que les couleurs ne peuvent pas correspondre à des températures.

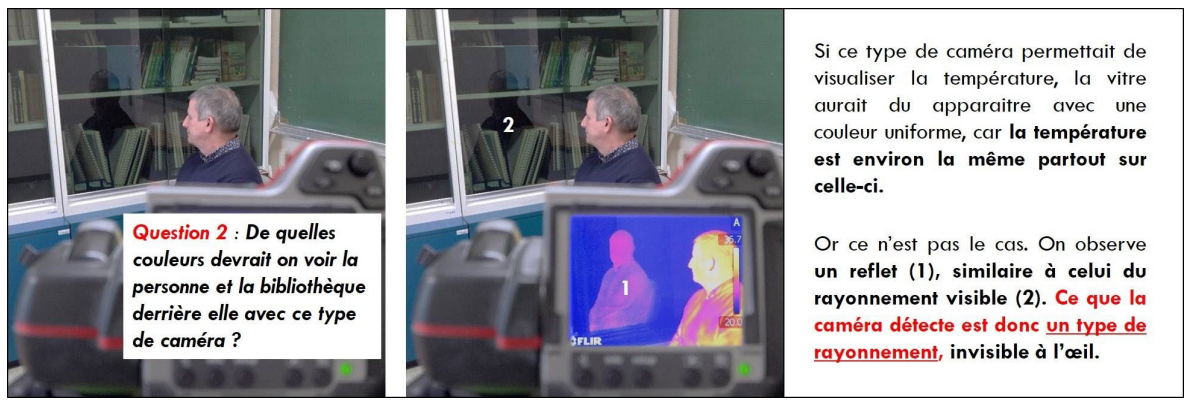

Figure n°3 – Mise en évidence du reflet infrarouge dans une vitre

Après avoir amené l'idée que la caméra permet de détecter un type de rayonnement, l'interprétation de la figure n°2 peut être clarifiée : en dessous d'une certaine température, le métal n'émet plus de rayonnement visible (dans le noir, on ne le voit pas). Il émet cependant toujours un rayonnement, invisible pour notre œil. Comme il s'agit d'un rayonnement émis après la zone rouge de plus en plus sombre, on l'appelle rayonnement infra-rouge. On privilégiera donc systématiquement l'appellation de « caméra infrarouge » plutôt que « caméra thermique ».

Reste à préciser la nature de la grandeur physique mesurée par ce détecteur. D'un point de vue physique, il s'agit de la puissance de rayonnement par unité de surface, pour une gamme de longueurs d'onde donnée. Le choix fait ici est de se restreindre à la notion de « puissance de rayonnement ». Le rapport à la surface est laissé implicite par le fait que les valeurs mesurées (traduites en couleurs) peuvent varier spatialement, ce qui sous-entend une grandeur locale sans nécessairement qu'il y ait besoin de l'expliciter la notion de puissance surfacique à ce stade. L'avantage de raisonner sur la notion de puissance est qu'elle peut s'appuyer sur l'idée intuitive de luminosité d'une lampe, ayant un sens à un instant donné.

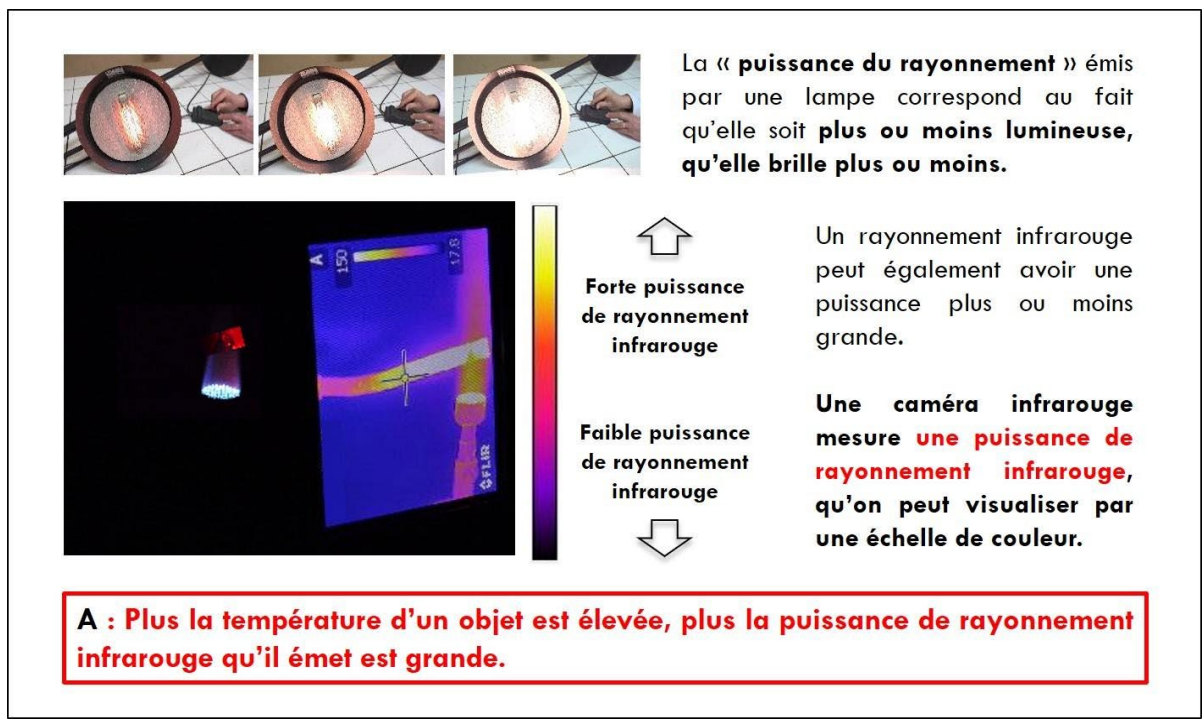

Figure n°4 : Introduction de la notion de puissance de rayonnement infrarouge

Une fois introduite la notion de puissance, le lien entre puissance de rayonnement émis et température peut être facilement mis en évidence. Dans le cas de la barre de cuivre chauffée, il suffirait de montrer avec un thermomètre que plus on s'éloigne de la flamme, plus la température diminue. De même, la puissance de rayonnement infrarouge émise par la barre décroit avec la distance à la flamme. L'exemple de la barre de cuivre incandescente pouvant donner l'impression que seul des objets très chauds peuvent émettre un rayonnement infrarouge, il peut être intéressant d'associer explicitement une puissance de rayonnement infrarouge à des objets à des températures proches de la température ambiante, comme par exemple dans la figure n°3 avec un corps humain, des habits, une vitre…

## Rayonnement à travers la matière solide

Avant d'introduire l'interaction entre rayonnement infrarouge et gaz, nous proposons de commencer par considérer le cas de solide (de même que dans (Dufresne, 2020; Thollot & Dequincey, 2021)). L'expérience de la figure n°5 a été pensé cependant de manière à servir d'appui pour interpréter celles avec le $CO_2$, comme nous le montrerons ensuite.

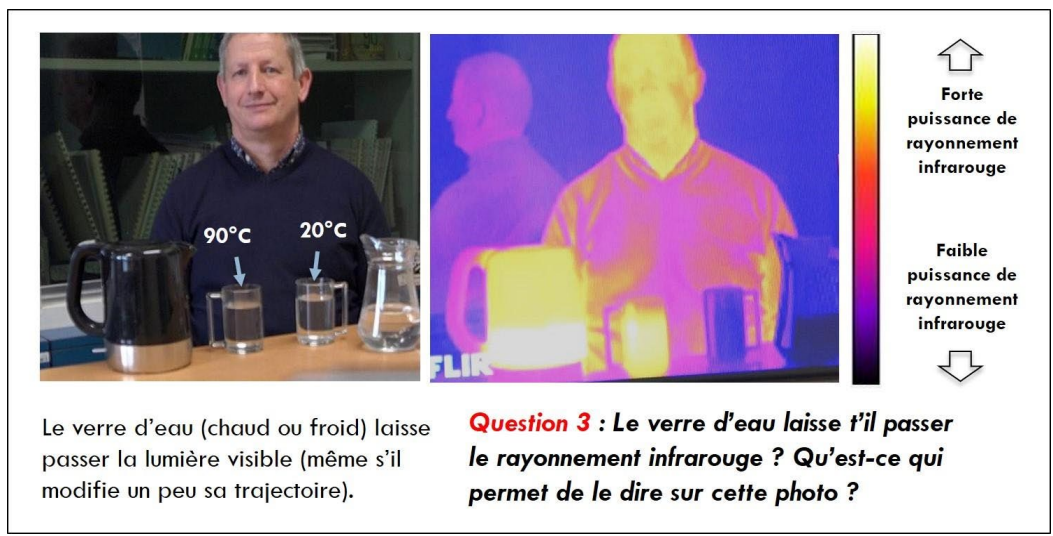

Figure n°5 : Rayonnement visible et infrarouge à travers la matière

Cette photo permet plusieurs observations :

- La puissance de rayonnement infrarouge émis par le verre d'eau chaude est plus grande que celle du verre d'eau froide.
- La puissance du rayonnement infrarouge en provenance du verre d'eau froide est plus faible que la puissance du rayonnement venant de la personne derrière. Le verre ne laisse donc pas passer (ou du moins pas entièrement) le rayonnement infrarouge. A ce stade, il n'est pas encore possible d'introduire la notion d'« absorption » du rayonnement par la matière, car cette baisse de puissance est également liée à la réflexion du rayonnement infrarouge par le verre.
- Dans le cas du verre d'eau chaude, même si une part du rayonnement venant de la personne ne traverse pas non plus le verre, cela ne peut pas se voir : la puissance du rayonnement émis par le verre chaud est suffisamment forte pour compenser la part venant de derrière qui ne traverse pas (et même être plus grande que celle-ci).

Un second exemple d'interaction rayonnement matière est également utilisé, celui d'une pochette plastique (figure n°6).

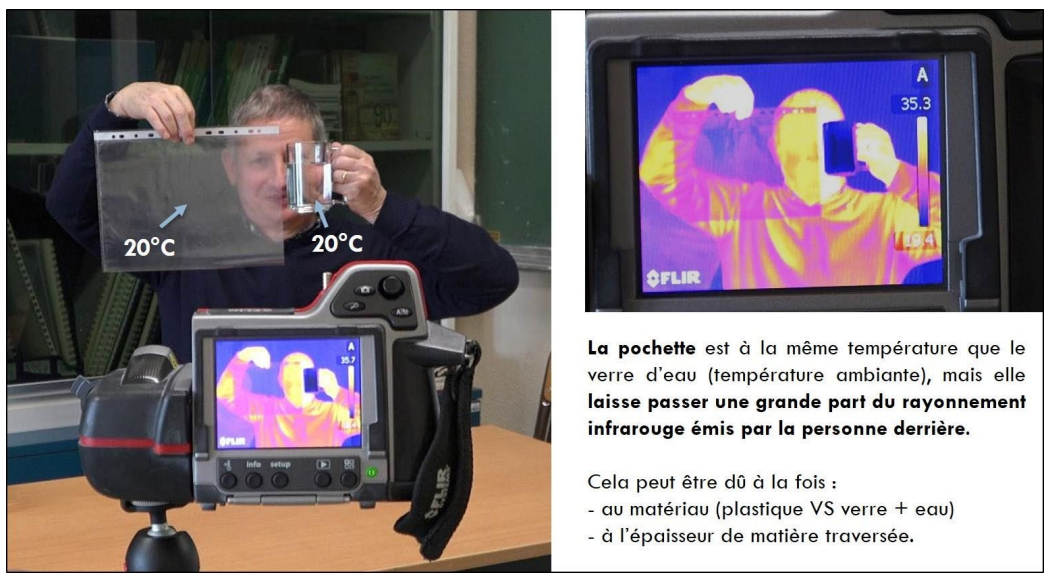

Figure n°6 : Rayonnement infrarouge à travers une pochette plastique et un verre d'eau

Après avoir considéré ces différents exemples, la question peut être posée en ce qui concerne les gaz. En effet, comme on l'a vu précédemment avec l'enquête effectuée à ce sujet, il semble que pour la large majorité des élèves du secondaire l'idée que les gaz puissent émettre ou absorber du rayonnement n'est pas une évidence.

### Rayonnement à travers les gaz

Les expériences qui suivent visent à mettre en évidence l'interaction entre $CO_2$ et rayonnement infrarouge. Malheureusement, elles ne sont pas possibles avec les caméras infrarouges les plus courantes (celles présentes dans les établissements scolaires) dont la gamme spectrale de sensibilité ne se superpose pas[10] avec celle du spectre d'absorption du $CO_2$. Ces expériences ont été réalisé avec une caméra infrarouge prêtée par un laboratoire de physique. Si l'on peut regretter l'impossibilité de reproduire ces expériences en classe, il se trouve que c'est également le cas des expériences permettant d'obtenir les spectres d'absorption des gaz. La grande différence ici est que l'interprétation des expériences que nous proposons est beaucoup plus

---

[10] La gamme spectrale des caméras infrarouge les plus courante se situe entre 8 et 15µm, ce qui correspond précisément à une plage pour laquelle le $CO_2$ et la vapeur d'eau n'émettent pas de rayonnement infrarouge. Ces caméras sont en effet conçues notamment pour pouvoir relier la puissance infrarouge provenant d'un bâtiment à son champ de température, sans que cela soit affecté par la présence de gaz entre le bâtiment et la caméra, $CO_2$ ou vapeur d'eau. Malheureusement, les caméras infrarouges permettant de détecter les gaz sont beaucoup plus chères (de l'ordre de 50 000 €).

accessibles, dans la mesure elle ne nécessite pas la notion de longueur d'onde électromagnétique.

La première expérience consiste comparer des ballons en latex, l'un rempli d'air, l'autre de $CO_2$ pur. A température ambiante, si l'on observe les deux ballons avec la caméra infrarouge, on observe la même chose pour les deux : une puissance uniforme. L'interprétation de chacun des ballons est cependant différente. Dans le cas du ballon d'air, une grande part du rayonnement infrarouge arrivant vers le ballon le traverse, et très peu de rayonnement est émis par celui-ci. Dans le cas du ballon de $CO_2$, une part du rayonnement arrivant vers le ballon est absorbé par le $CO_2$, et celui émis également du rayonnement infrarouge. Ainsi, pour comprendre cette interprétation, il faut déjà pouvoir mettre en évidence l'absorption et l'émission du $CO_2$. Une manière de le faire consiste à observer les ballons à plus basse ou plus haute température que leur environnement, de façon à ce que la puissance de rayonnement émis ne compense pas la puissance de rayonnement absorbée. Nous commencerons avec des ballons refroidis dans un bac à glace (figure n°7).

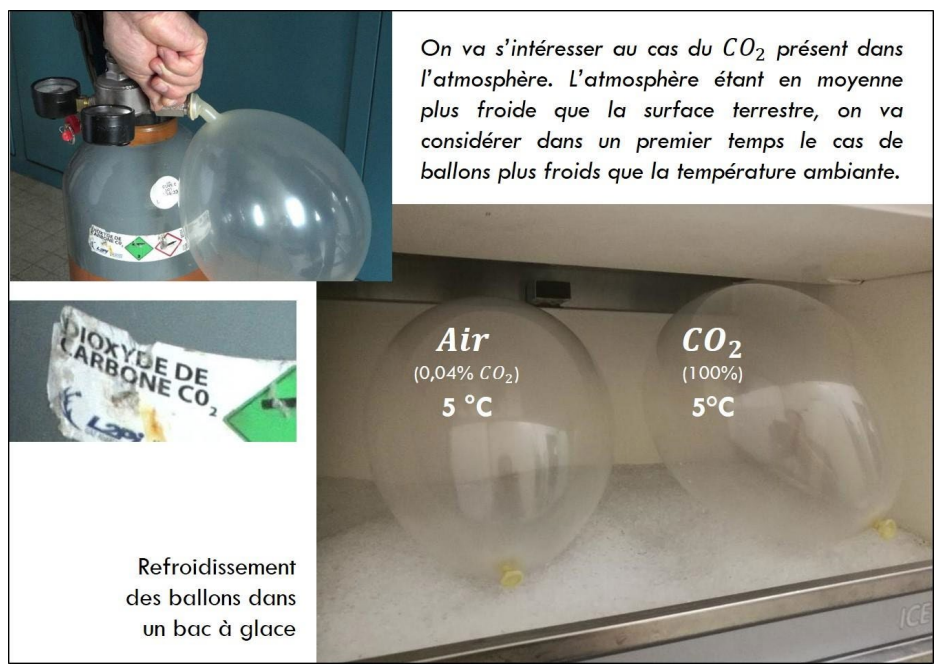

Figure n°7 : Gonflage d'un ballon de $CO_2$ et refroidissement des ballons

En lumière visible, on constate dans un premier temps que le $CO_2$ est aussi transparent que l'air. Observé avec la caméra infrarouge, on obtient l'image suivante (figure n°8) :

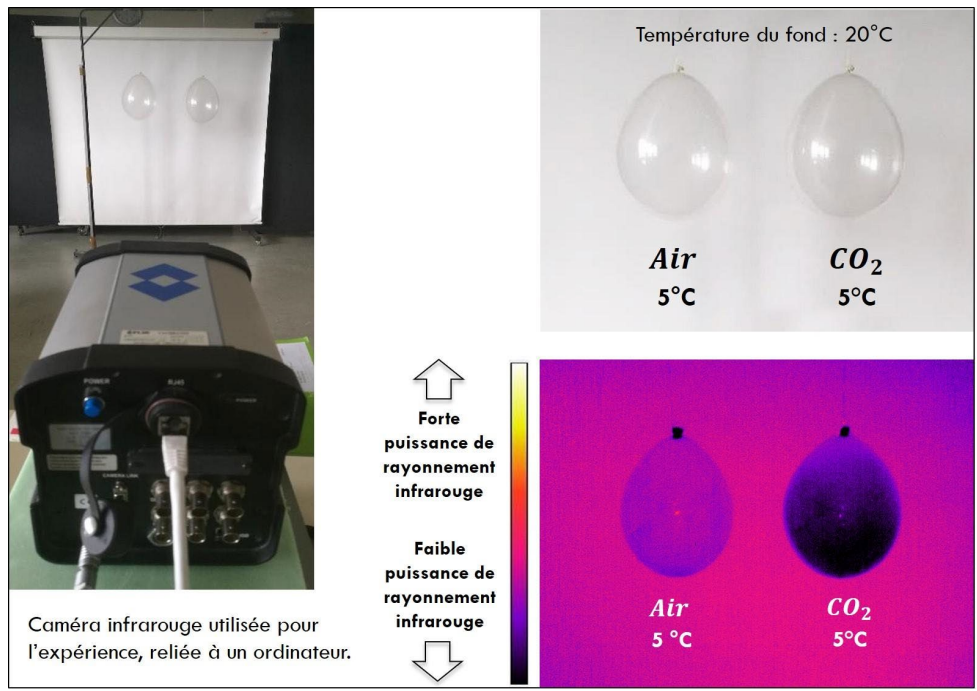

Figure n°8 : Ballons d'air et de $CO_2$ froids observés avec une caméra infrarouge

La comparaison avec le ballon d'air à la même température permet de dissocier l'influence du $CO_2$ de celle de la membrane du ballon, qui à la fois peut réfléchir et absorber une part du rayonnement infrarouge. La seconde expérience consiste à observer des ballons plus chauds que la température ambiante (figure n°9).

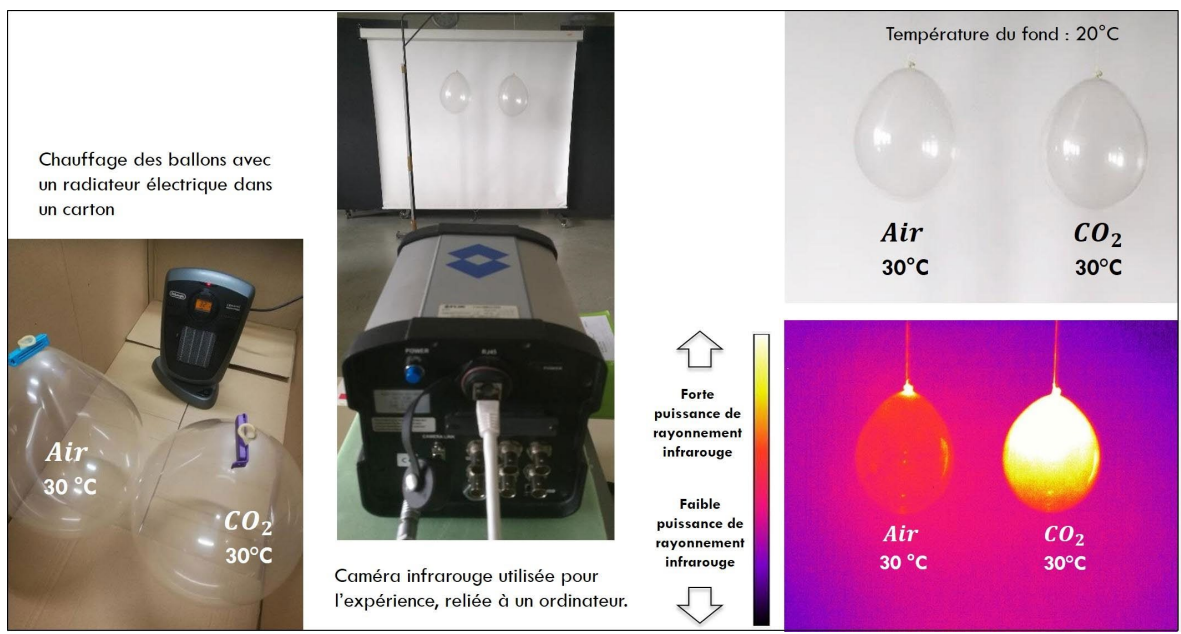

Figure n°9 : Ballons d'air et de $CO_2$ chauds observés avec une caméra infrarouge

La comparaison complète peut alors être faite avec le cas des verres d'eau chaud et froid (figure n°10).

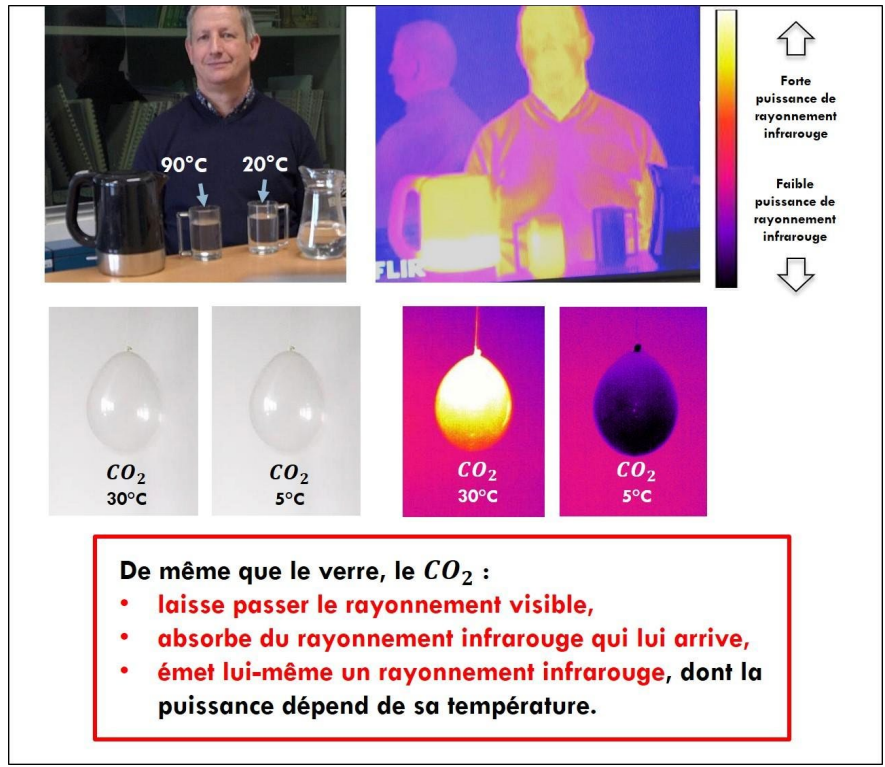

Figure n°10 : Comparaison entre le verre et le $CO_2$

L'extrapolation au système Terre-atmosphère peut alors être questionné (figure n°11) et établie (figure n°12).

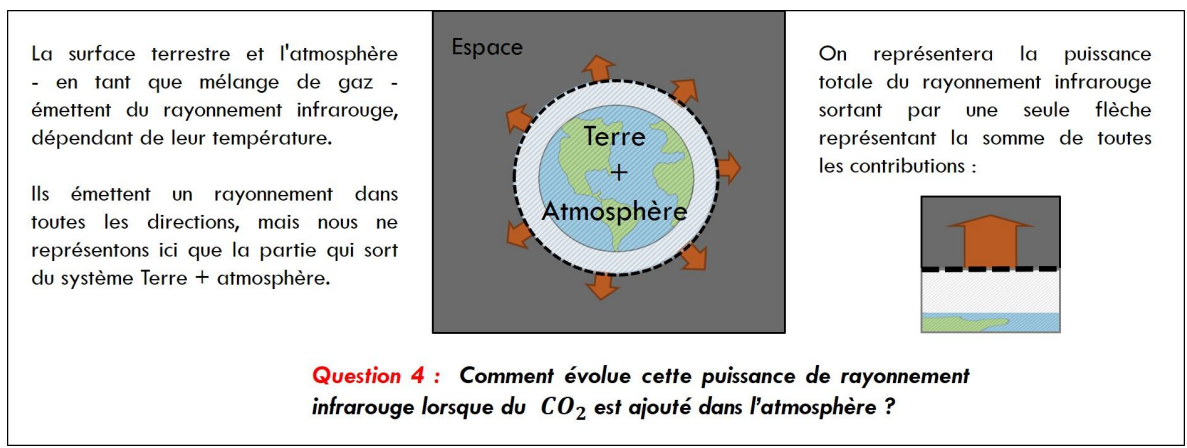

Figure n°11 : Deux représentations du rayonnement infrarouge émis par le système Terre-atmosphère

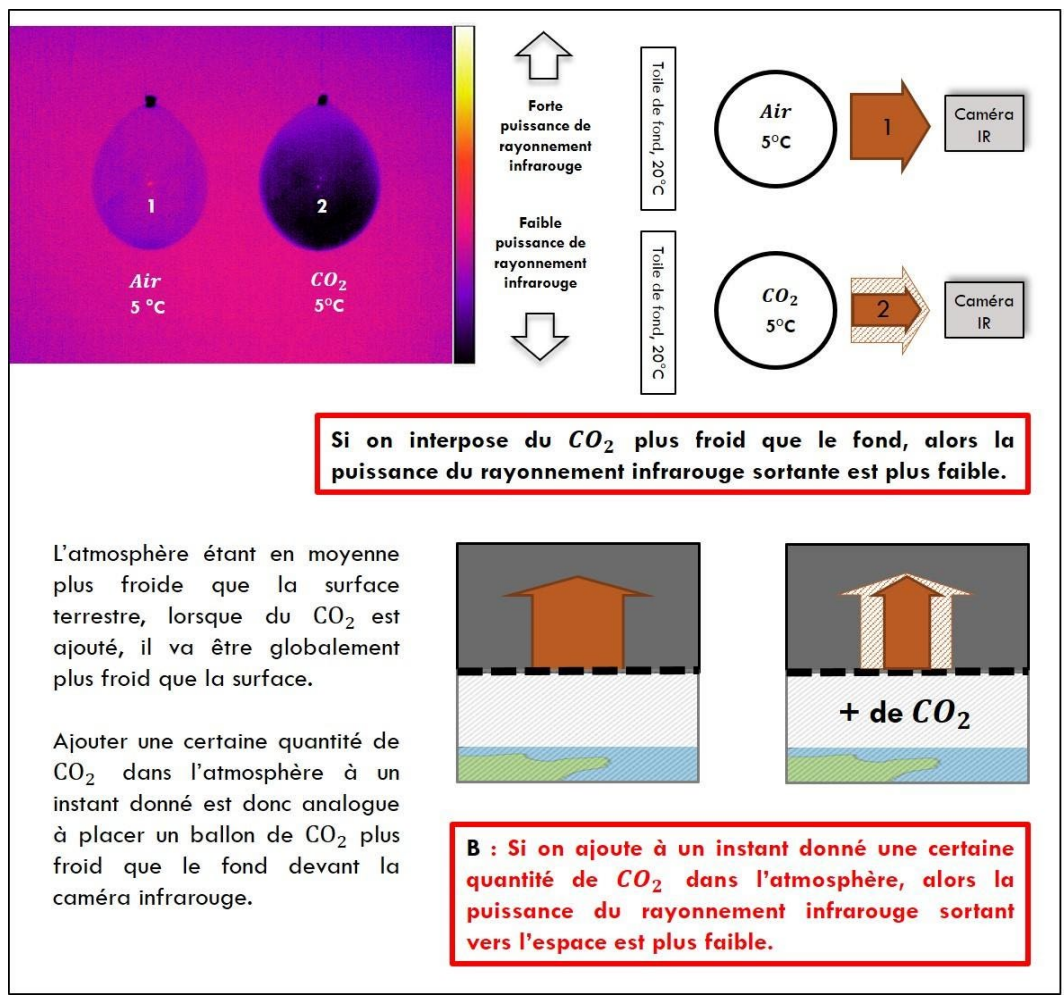

Figure n°12 : Extrapolation de la conséquence d'un ajout de $CO_2$ dans l'atmosphère sur le rayonnement infrarouge sortant vers l'Espace

## Bilan de rayonnements et évolution de la température

Le dernier élément nécessaire pour relier $CO_2$ et température est la notion de bilan de rayonnements. L'exemple choisi est celui d'une plaque de tôle au soleil (figure n°13).

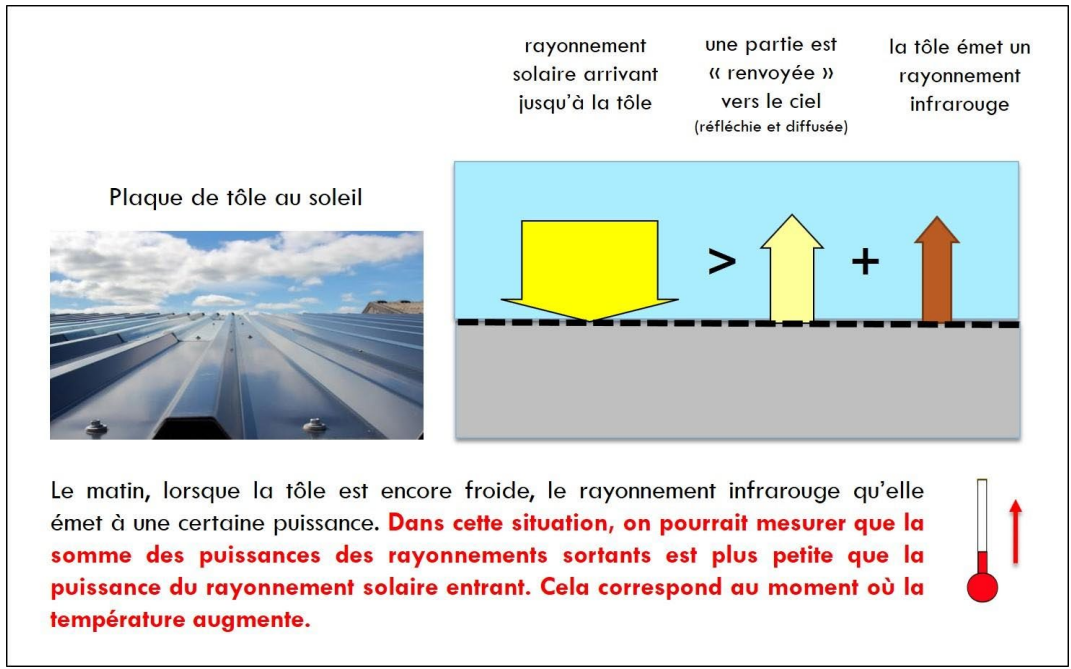

Figure n°13 : Bilan des rayonnements pour une plaque de tôle au Soleil

Cet exemple permet de relier la comparaison des puissances de rayonnement entrant et sortant à l'évolution de la température (figure n°15).

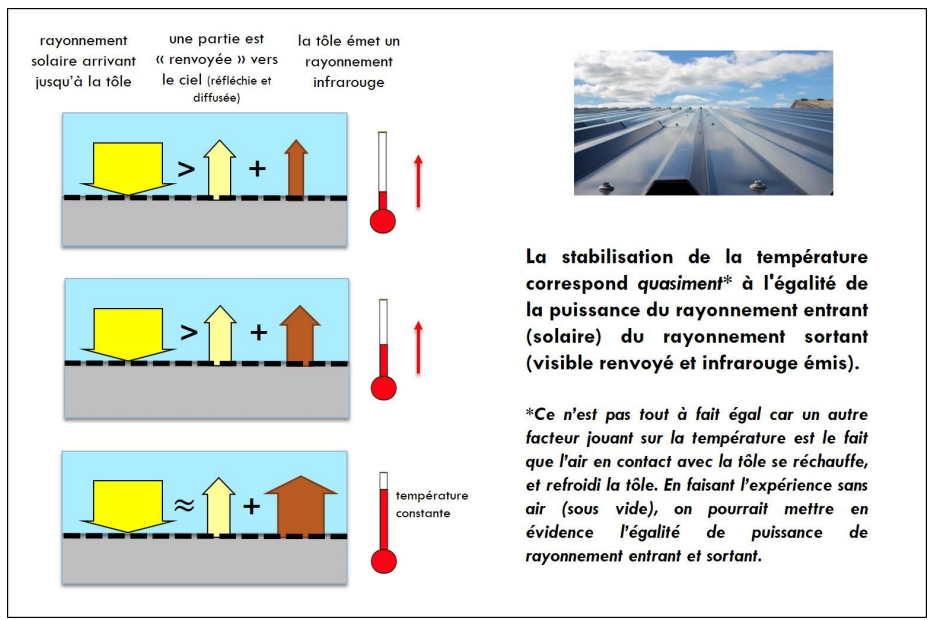

Figure n°14 : Bilan des rayonnements et évolution de la température, dans l'air

La prise en compte de l'influence de l'air sur la température de tôle mène à imaginer le cas d'une plaque de métal dans une enceinte sous vide. Cette expérience va correspondre à la situation du système Terre-atmosphère, dans le vide de l'Espace.

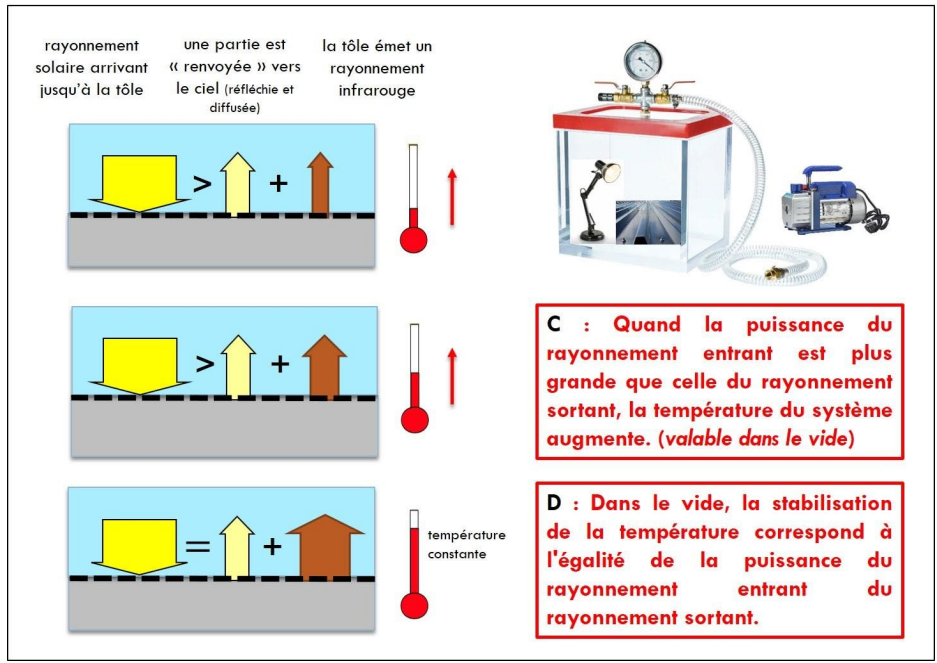

Figure n°15 : Bilan des rayonnements et évolution de la température, dans le vide

**Bilan des rayonnements du système Terre-atmosphère**

La dernière étape du cheminement logique proposé consiste à mettre en relation les éléments des parties 4 et 5, dans le cas du système Terre-atmosphère (figure 16).

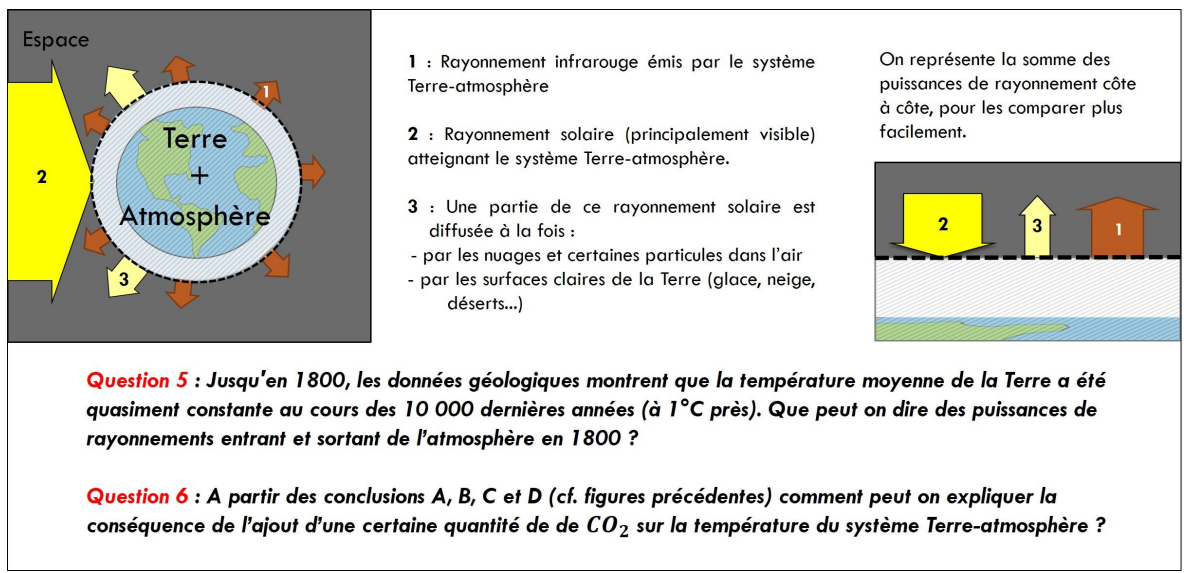

Figure n°16 : Bilan des rayonnements du système Terre-atmosphère

Ici, un choix didactique central consiste à ne pas représenter les différents rayonnements émis par la surface terrestre et l'atmosphère, pour se concentrer uniquement sur l'interface entre le système Terre-atmosphère et l'Espace. La raison de ce choix et qu'une représentation du bilan des rayonnements impliquant la surface terrestre et l'atmosphère (figure n°17) ne permet pas d'aller au bout de raisonnement suite à l'ajout de $CO_2$.

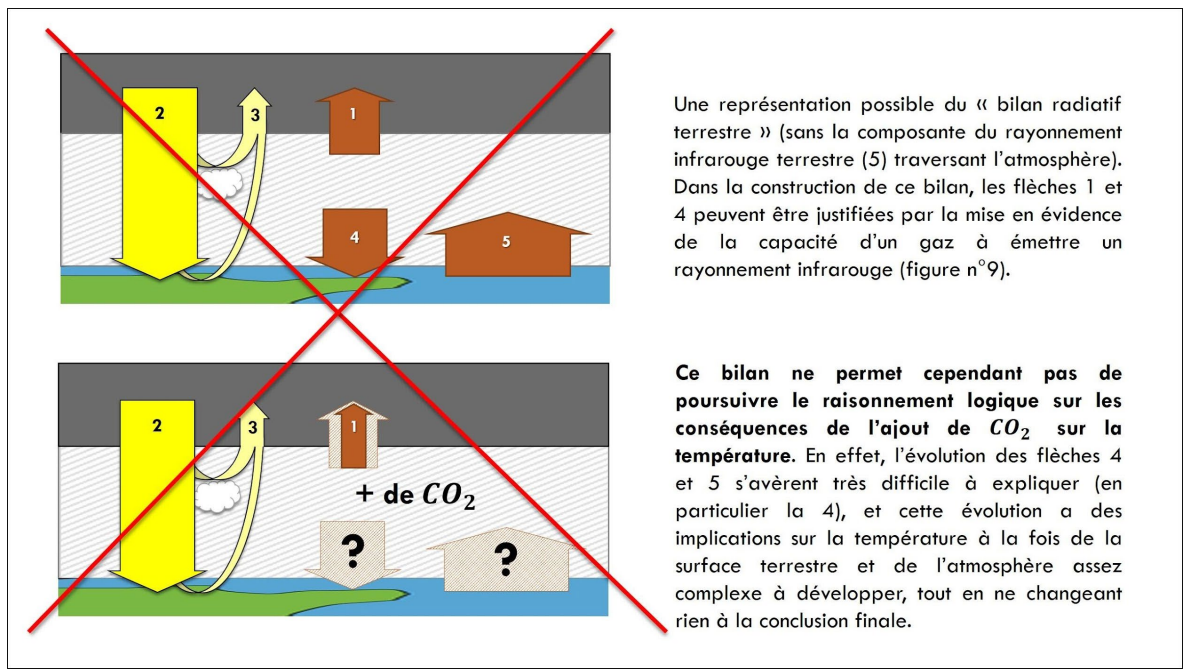

Figure n°17 : Bilan radiatif prenant en compte les interfaces Terre-Atmosphère et Atmosphère-Espace (non utilisé dans la proposition présentée ici)

En effet, ce bilan ferait figurer notamment le rayonnement infrarouge émis par l'atmosphère vers la Terre (4 dans la figure n°17). Or celui-ci est également affecté par l'ajout de $CO_2$, d'une manière contre-intuitive et difficile à expliquer pour un public non spécialiste. Ce niveau de détail n'est par ailleurs pas nécessaire pour arriver à la conclusion recherchée, pour laquelle un raisonnement sur le schéma de la figure n°16 suffit, comme par exemple celui de la figure n°18, répondant aux questions 5 et 6 de la figure n°16.

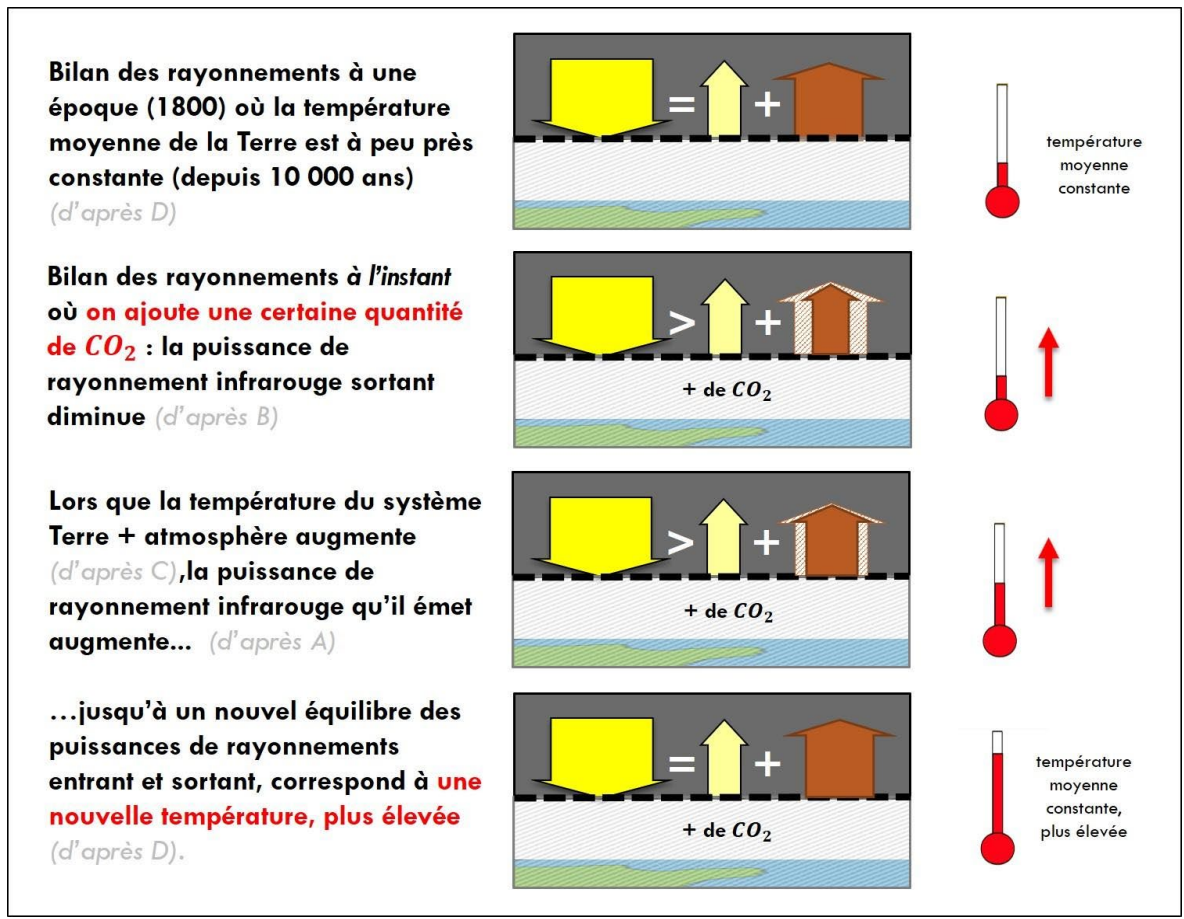

Figure n°18 : Conséquence de l'ajout de $CO_2$ sur l'évolution de la température du système Terre-atmosphère

Dans le cadre de cette représentation, il est également possible de considérer les effets amplificateurs du réchauffement, comme par exemple avec la figure n°19.

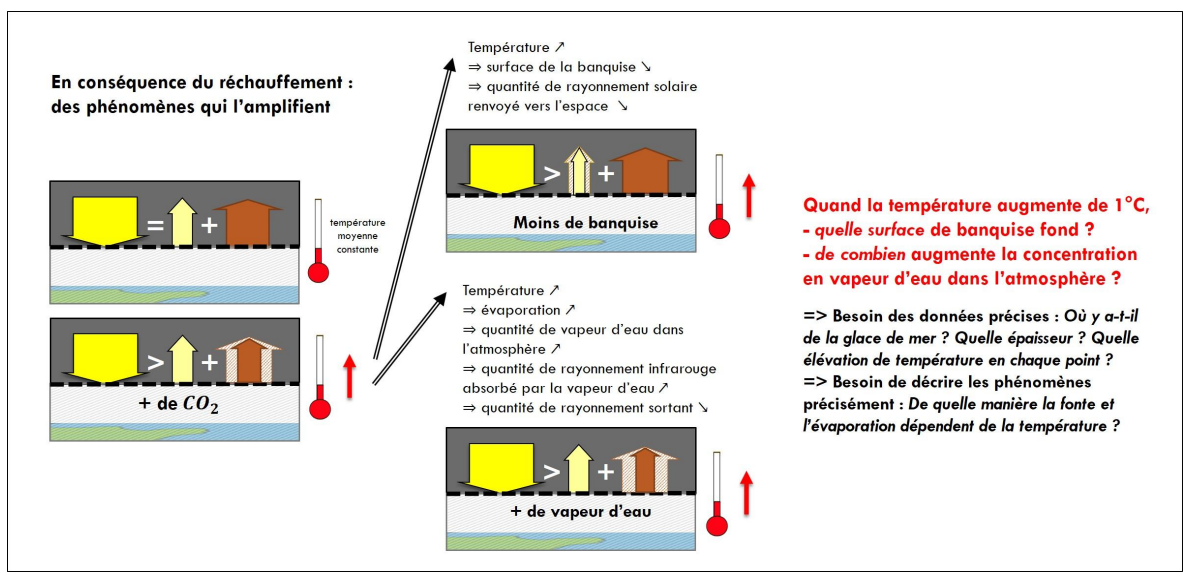

Figure n°19 : Deux phénomènes d'amplification du réchauffement

### Forçage radiatif

Enfin, l'approche proposée permet également une clarification et représentation de la notion de « forçage radiatif » associé à un scénario du GIEC, tel que proposé dans la figure n°20. En effet, la définition du programme d'Enseignement Scientifique de Terminale (« différence entre l'énergie radiative reçue et l'énergie radiative émise »), reprise par les manuels scolaires, est en contradiction avec l'usage qui en est fait dans les documents pédagogiques associés. Dans un contexte où les bilans radiatifs sont toujours effectués à un instant donné, la définition du programme évoque en effet celle du « forçage radiatif instantané ». Or les nombres 8,5 / 7,0 / 4,5 / 2,6 / 1,9 W/m², présentés comme caractérisant les différents scénarios du GIEC via leur « forçage radiatif en 2100 », ne correspondent pas à définition précédente, mais à celle introduite dans la figure n°20.

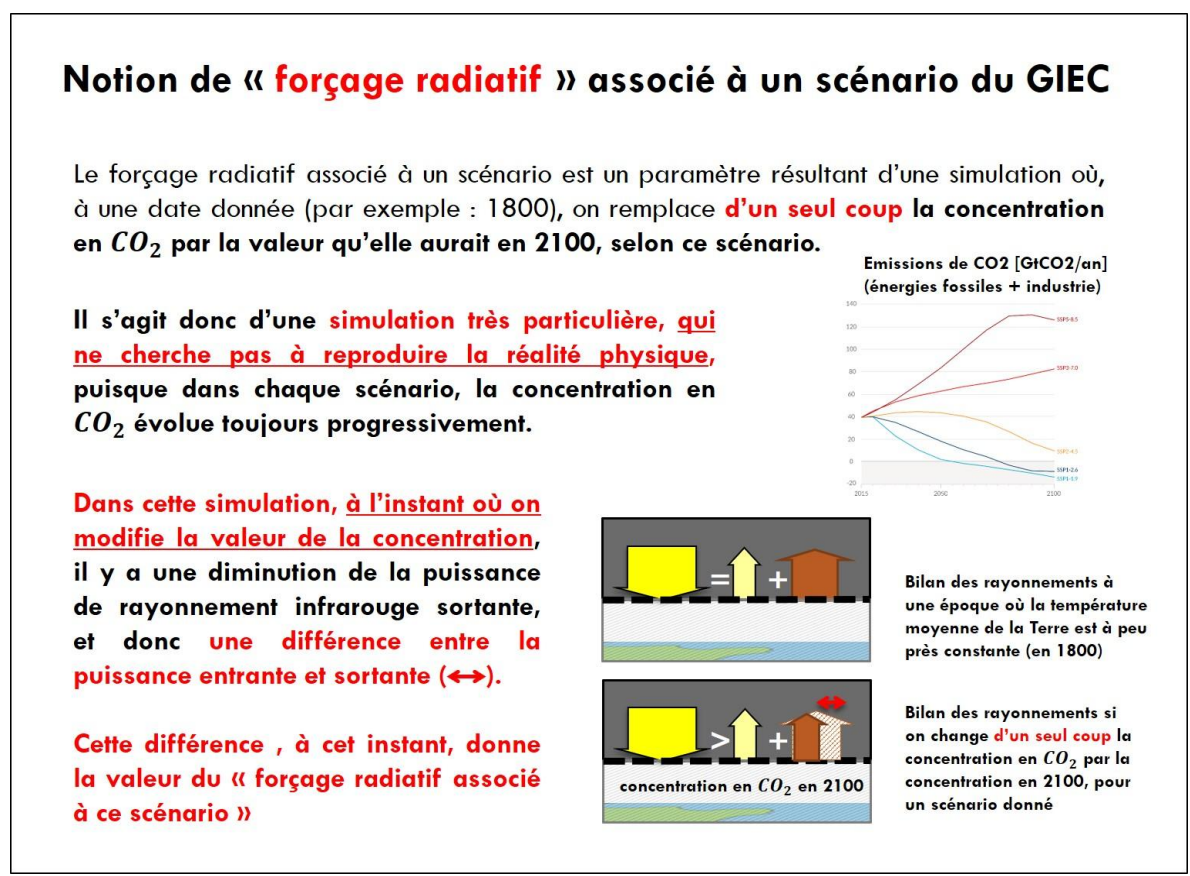

Figure n°20 : Clarification de la définition du forçage radiatif associé au scénarios du GIEC

# Conclusion

### Le cœur de l'approche proposée

L'approche proposée vise à construire le lien entre émissions de $CO_2$ dans l'atmosphère et température moyenne de la Terre, pour un public non spécialiste de physique. Elle s'articule autour de deux points centraux, à notre connaissance inédits dans les approches existantes sur le sujet :

- l'interprétation d'expériences réalisées avec une caméra infrarouge d'une gamme spectrale permettant de voir l'émission et l'absorption du $CO_2$,
- le raisonnement sur un bilan radiatif à l'interface entre le système Terre-atmosphère et l'Espace, sans considération de ce qui se passe entre la surface terrestre et l'atmosphère.

### Minimalisme conceptuel

Ce dernier point rejoint l'intention de produire une proposition aussi accessible que possible, c'est à dire avec le minimum de prérequis conceptuels. Cela a mené en particulier à construire un cheminement logique ne s'appuyant à aucun moment sur les concepts physiques suivants :

- **La notion d'onde électromagnétique, de longueur d'onde et de spectre de la lumière.** Si ces notions peuvent être évoquées dans le cadre du programme de seconde de physique-chimie, elle reste d'une abstraction extrêmement élevée pour des élèves ne pouvant pas avoir d'idée à ce niveau de ce que représente les champs électrique et magnétique. De plus, l'affirmation selon laquelle « la lumière est une onde électromagnétique » relève au lycée (et même souvent au-delà), du pur argument d'autorité. Raisonner à partir des longueurs d'onde reviendrait alors à prolonger cet argument d'autorité, ce qui aurait été contraire à l'intention de construire une connaissance, au cœur de notre démarche.
- **La notion d'énergie**. Bien que présente tout au long du cursus scolaire, en physique-chimie comme dans d'autres disciplines, les recherches en didactique et en épistémologie montrent l'importante confusion associée à l'apprentissage du concept d'énergie, faisant écho à sa très grande abstraction (voir par exemple la synthèse effectuée dans (Bächtold, 2018), chapitre 3). Plutôt que de considérer le « bilan énergétique terrestre », nous n'avons considéré que des « bilans de puissances de rayonnement ». Si le contexte s'y prête, l'introduction de la notion d'énergie est toujours possible *a posteriori*, en tant que produit de la puissance de rayonnement par une durée.

Nous avons également fait le choix de ne pas nous appuyer du tout sur d'autres concepts de physique couramment utilisés dans les approches du sujet. Il s'agit par exemple des notions suivantes de flux, chaleur, transfert thermique, équilibre dynamique, régime transitoire et permanent.

Cette posture minimaliste reprend celle défendue dans (Maron, 2015) :

> « le critère de restriction au minimum nécessaire est lié à la nature du contenu en jeu : étant donnée la complexité [de celui-ci], développer des éléments qui n'y sont pas directement reliés reviendrait à diluer l'articulation des idées dans une quantité de connaissance plus grande encore. Il s'agit donc d'appliquer un principe de parcimonie pour la sélection des éléments à considérer (également connu sous le nom de rasoir d'Ockham : « Les entités ne doivent pas être multipliées par-delà ce qui est nécessaire ») »

Ce travail de montre qu'il est possible de construire une approche qualitative rigoureuse pour relier émissions de $CO_2$ et réchauffement climatique, sans avoir besoin de faire appel à un grand nombre de notions abstraites inaccessibles à un public non spécialiste ou débutant en physique. La prise en compte de phénomènes d'amplification du réchauffement (ou « rétroaction positive ») permet de mettre en évidence la nécessité d'une description locale du système et la quantification des différents phénomènes en jeu (questions de la figure n°19, à droite). Autrement dit, elle mène à un travail sur la construction d'un modèle de climat, dont une approche didactique sera proposée dans un article à venir, avec la même intention d'être aussi accessible que possible.

## Remerciements

## Références